\documentclass[%
reprint,
superscriptaddress,
amsmath,amssymb,
prl,
]{revtex4-2}

\usepackage{color}
\usepackage{csquotes}

\usepackage{ulem}
\usepackage{graphicx}
\usepackage{dcolumn}
\usepackage{bm}
\usepackage[unicode=true,colorlinks=true,citecolor=blue,urlcolor=blue]{hyperref}

\newcommand{\eps}{\varepsilon}
\newcommand{\e}{\mathrm{e}}
\renewcommand{\i}{{\rm i}}
\renewcommand{\d}{\mathrm d}
\renewcommand{\Re}{\mathop{\rm Re}}
\renewcommand{\Im}{\mathop{\rm Im}}
\usepackage{braket}

\begin{document}

\author{V.~O.\ Kozlov}
\affiliation{Spin Optics Laboratory, St.\,Petersburg State University, 198504 St.\,Petersburg, Russia}
\affiliation{Photonics Department, St.\,Petersburg State University, Peterhof, 198504 St.\,Petersburg, Russia}

\author{N.~S.\ Kuznetsov}
\affiliation{Spin Optics Laboratory, St.\,Petersburg State University, 198504 St.\,Petersburg, Russia}
\affiliation{Photonics Department, St.\,Petersburg State University, Peterhof, 198504 St.\,Petersburg, Russia}

\author{D.~S.\ Smirnov}
\affiliation{Ioffe Institute, St.\,Petersburg, Russia}
\affiliation{Spin Optics Laboratory, St.\,Petersburg State University, 198504 St.\,Petersburg, Russia}

\author{I.~I.\ Ryzhov}
\affiliation{Photonics Department, St.\,Petersburg State University, Peterhof, 198504 St.\,Petersburg, Russia}
\affiliation{Spin Optics Laboratory, St.\,Petersburg State University, 198504 St.\,Petersburg, Russia}

\author{G.~G.\ Kozlov}
\affiliation{Spin Optics Laboratory, St.\,Petersburg State University, 198504 St.\,Petersburg, Russia}
\affiliation{Solid State Physics Department, St.\,Petersburg State University, Peterhof, 198504 St.\,Petersburg, Russia}

\author{V.~S.\ Zapasskii}
\affiliation{Spin Optics Laboratory, St.\,Petersburg State University, 198504 St.\,Petersburg, Russia}

\begin{abstract} 

{It is known that linear birefringence of the medium essentially hinders measuring the Faraday
effect. For this reason, optically anisotropic materials have never been considered as objects of the
Faraday-rotation-based spin noise spectroscopy (SNS). We show, both theoretically and experimentally, that strong optical anisotropy that may badly suppress the regular Faraday rotation of the medium, practically does not affect the measurement of the spatially uncorrelated spin fluctuations.
An important consequence of this result is that the Faraday-rotation noise should be also insensitive to spatially nonuniform birefringence, which makes the SNS applicable to a wide class of optically anisotropic and inhomogeneous materials. We also show that the birefringent media provide additional opportunity to measure spatial spin correlations. Results of the experimental
measurements of the spin-noise spectra performed on 4$f$--4$f$ transitions of Nd$^{3+}$ ions in the CaWO$_4$ and LiYF$_4$ crystals well agree with the theory. }

\end{abstract}

\title {Spin noise in birefringent and inhomogeneous media}
\maketitle


{\it Introduction.}
In the past decade, the spin noise spectroscopy (SNS) has gained an important place in the experimental physics of magnetic resonance. Being primarily demonstrated on atomic systems \cite{zap-frns81}, this technique is, at present, widely used in studies of electron and nuclear spin systems in semiconductors~\cite{muller-semicon-sns10,crooker-sns-e-holes-qd2010,dahbashi-sn-single-hole14,berski-ultrahigh-bw-sns13,zapasskii-sns-review13,zapasskii-optical-sns13,glasenapp-sns-beyond-thermal-equil14,hubner-sns-semicond14,berski-e-nucl-sn15,cronenberger-spatiotemporal-sns2019,smirnov-sn-review2021} and has been recently successfully applied to dielectric crystals with paramagnetic impurities \cite{kamenskii-re-sns2020,kamenskii-invariants21}. As compared with the conventional EPR spectroscopy, the SNS offers a number of new specific possibilities of research and reveals many features more typical for nonlinear optics~\cite{yang-two-color-sns14,romer-spatially-resolved-sns09, zapasskii-optical-sns13,fomin-high-spin21,fomin-anomal-broad-sn-cs21,glazov-linear-optics-raman-sns2015}.
 
At the same time, it might seem that the SNS, based on detection of the Faraday rotation (FR) noise, should have inherited the main features (both merits and drawbacks) of the FR method. Specifically, the FR measurements are known to be highly sensitive to linear birefringence of the medium~\cite{nakagawa-faraday19,ramachandran-mo-rot-bf52,ramaseshan-faraday-eff-bf51,woodford-interpr-magn-fr07}. This circumstance, for a long time, prevented application of the SNS to optically anisotropic crystals. In this Letter, we show that linear birefringence, generally, has no significant  effect on the stochastic FR noise.

The fact that the linear birefringence affects the spatially uniform and spatially fluctuating FR (gyration) in a strongly different way, can be supported by the following reasoning. 
Polarization of the light travelling through a birefringent medium exhibits spatial oscillations, with the $\sigma_+$ and $\sigma_-$ components being periodically interchanged. As a result, the sign of the FR is periodically inverted so that the total FR appears to be close to zero. 
For the spatially uncorrelated gyration, this mechanism of the FR suppression does not work, because any inversion of the random FR leaves it random, and the measured FR noise remains the same.

{In this work, we present a rigorous theoretical description of the effect of linear birefringence upon the FR noise detected in the SNS experiments. The results of this treatment allow us to make positive conclusions about applicability of the SNS to spatially inhomogeneous optically anisotropic media. We also show that the FR noise signal, in a uniformly birefringent medium, may be enhanced or suppressed for spin arrangement spatially correlated at the wave vector of the intrinsic polarization beats of the birefringent medium. Applicability of the SNS to birefringent media is illustrated by experimental study of spin noise in rare-earth-doped crystals. }

\vspace{10pt}
{\it Theory}. We consider the light propagation along one of the principal axes ($z$) of a birefringent crystal, with two other principal axes $x$ and $y$ being characterized by the dielectric constants $\eps_x$ and $\eps_y$, respectively. The case of the light propagation in arbitrary direction can be considered in a similar way (see {\it Discussion} below). The Fourier component of electric field of the probe light at the frequency $\omega$ is given by
\begin{equation}
 E_{x,y}^{(0)}(z)=E_{x,y}^{(0)}(0)\exp(\i k_{x,y}z)
\end{equation}
where $k_{x,y}=k_0\sqrt{\eps_{x,y}}$ are the wave vectors of the light components polarized along the corresponding axes, $k_0=\omega/c$, and the sample face is located at $z=0$.

The spin fluctuations in the crystal provide a stochastic polarizability with the off-diagonal components $\chi(z)=\chi_{xy}=-\chi_{yx}$, which are odd under time reversal~\cite{landau-8-electrodyn84}. It is also a random function of the $x$ and $y$ coordinates, but this dependence is inessential for the calculation of the FR noise spectra~\cite{smirnov-sn-review2021}, so we ignore it. The stochastic polarizability produces the following additional polarization in the sample:
\begin{equation}
 \Pi_{x,y}(z)=\pm\chi(z)E_{y,x}^{(0)}(0)\e^{\i k_{y,x} z},
\end{equation}
The polarization, in turn, produces scatted field, which at the face of the sample of the length $L$ takes the form~\cite{yugova09}
\begin{equation}
 \label{eq:E1}
 E_{x,y}^{(1)}(L)=\pm\frac{2\pi\i k_0^2}{k}E_{y,x}^{(0)}(L)\chi_\pm,
\end{equation}
where we assume $|k_x-k_y|\ll k$, $k=(k_x+k_y)/2$, and
\begin{equation}
 \label{eq:chi_def}
 \chi_\pm=\int\limits_0^L\d z\chi(z)\e^{\pm\i(k_x-k_y)(L-z)}
\end{equation}
are the two spatial harmonics of the fluctuating polarizability induced by the spin noise.

\begin{figure}[t]
\includegraphics[width=0.75\linewidth]{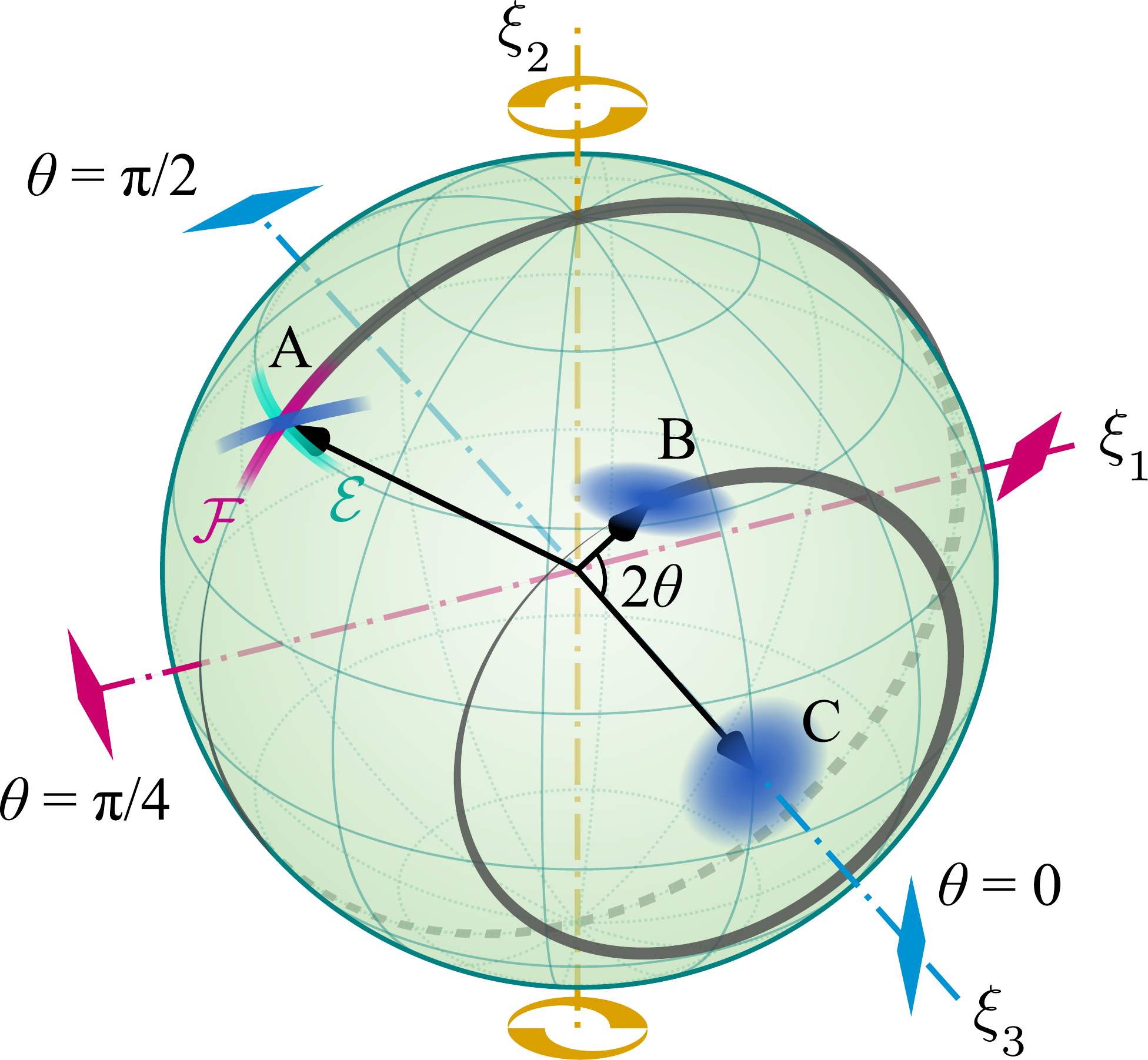}
\caption{Schematic presentation of the Stokes vector fluctuations for the light transmitted through a birefringent crystal. The axis $\xi_3$ corresponds to the polarizations along the main axes.
The points $A$, $B$, and $C$, on the surface of the Poincar\'e sphere correspond to azimuths of the incident light polarization {$\theta$} equal to $\pi/4$, approximately $\pi/8$, and 0. Conversion of the arc segment at $A$ to the round spot at $C$ indicates transformation of uncorrelated FR and ellipticity noises to correlated ones.}
\label{fig:poincare}
\end{figure}

Generally, the linearly polarized light, after passing through the birefringent medium, becomes elliptically polarized. The polarization is described by the Stokes parameters $\xi_1$, $\xi_2$, and $\xi_3$, which fluctuate due to the spin noise in the sample. We define the generalized FR and ellipticity signals as~\cite{yugova09}
\begin{equation}
 \label{eq:FE_def}
 \mathcal F=\frac{\braket{\xi_3}\delta\xi_1-\braket{\xi_1}\delta\xi_3}{\sqrt{\braket{\xi_1}^2+\braket{\xi_3}^2}},
 \quad
 \mathcal E=\frac{\delta\xi_2}{\sqrt{1-\braket{\xi_2}^2}},
\end{equation}
respectively, where $\braket{\xi_i}$ with $i=1,2,3$ are the average Stokes parameters and $\delta\xi_i=\xi_i-\braket{\xi_i}$ are their deviations proportional to $E_{x,y}^{(1)}$ and to the stochastic spin polarization. Generally, on the Poincar\'e sphere, $\mathcal F$ and $\mathcal E$ represent fluctuations of the Stokes vector in the equatorial and meridional directions, respectively, as shown schematically in Fig.~\ref{fig:poincare} at the point $A$.

To calculate the noise intensities of $\mathcal F$ and $\mathcal E$, we neglect the spatial correlations in the fluctuating $\chi(z)$, as discussed above. It is convenient to introduce a reduced real-valued spin polarization in the sample $m(z)$ according to $\chi(z)=X m(z)$, where $X$ is a complex coefficient and the normalization condition for $m(z)$ reads
\begin{equation}
 \label{eq:m_def}
 \braket{m(z)m(z')}=\delta(z-z').
\end{equation}
Typically, at the optical resonances $X\propto1/[\gamma+\i(\omega_0-\omega)]$, where $\omega_0$ is the resonance frequency and $\gamma$ is its width~\cite{smirnov-sn-review2021}. Generally, one can directly express the intensities of the noise $\braket{\mathcal F^2}$ and $\braket{\mathcal E^2}$ through the coefficient $X$ and the dimensionless parameter $(k_y-k_x)L$.

Let us consider the limit of strong anisotropy, $q=k_x-k_y\gg 1/L$, assuming the absorption anisotropy (imaginary part of $k_x-k_y$) to be zero. From the definitions~\eqref{eq:chi_def} and~\eqref{eq:m_def} we obtain the following correlators of the polarizabilities: $\braket{\chi_+\chi_{-}}=X^2L$, $\braket{\chi_+^*\chi_{-}^*}=X^{*2}L$, $\braket{\chi_+\chi_{+}^*}=\braket{\chi_{-}\chi_{-}^*}=|X^{2}|L$, all the other correlators between $\chi_\pm$ and $\chi_\pm^*$ are zero. Substituting these expressions in Eqs.~\eqref{eq:E1} and~\eqref{eq:FE_def} we obtain the total noise of the FR and ellipticity:
\begin{equation}
\label{eq:total_noise}
 \braket{\mathcal F^2}+\braket{\mathcal E^2}=\frac{1}{2}\mathcal Q|X|^2L\left[3+\cos(4\theta)\right],
\end{equation}
where $\mathcal Q=2\left(2\pi k_0^2/k\right)^2$ and $\theta$ is the angle between the incident light polarization plane and the $x$ axis.

{In the opposite limit of negligible birefringence, we obtain in the same way} $\braket{\mathcal F^2}+\braket{\mathcal E^2}=2\mathcal Q|X|^2L$. Thus, the birefringence suppresses the total polarization noise only $1$---$2$ times in contrast with the regular Faraday effect suppressed by the factor $qL\gg1$.

For the incident light polarized at $\theta=\pi/4$, we have $\braket{\mathcal F^2}=\mathcal Q(\Im X)^2L$ and $\braket{\mathcal E^2}=\mathcal Q(\Re X)^2L$, so that the FR and ellipticity noises are proportional to the regular FR and ellipticity in isotropic medium squared. Moreover, they are completely correlated, $\braket{\mathcal F\mathcal E}^2=\braket{\mathcal F^2}\braket{\mathcal E^2}$. On the Poincar\'e sphere, the FR and ellipticity fluctuations represent an arc, as shown in Fig.~\ref{fig:poincare} at the point $A$. These results are standard for the optical spin noise spectroscopy~\cite{smirnov-sn-review2021}.

By contrast, if the probe beam is initially polarized along one of the principal axes (e.g., $\theta=0$), the FR and ellipticity noise intensities are equal, $\braket{\mathcal F^2}=\braket{\mathcal E^2}$, and they are uncorrelated, $\braket{\mathcal F\mathcal E}=0$. This is illustrated by the blue {round spot} $C$ in Fig.~\ref{fig:poincare}. 
This surprising result is related to the presence of the two spatial harmonics of the spin noise at the wave vectors $\pm q$, as defined by Eq.~\eqref{eq:chi_def}.

{For the intermediate cases of the light polarization azimuth ($0<\theta<\pi/4$), the polarization noise on the Poincar\'e sphere represents an elliptical spot smoothly transforming from the arc to the circle, as shown in Fig.~\ref{fig:poincare} at the point $B$. The total intensity of the FR and ellipticity noises is generally described by Eq.~\eqref{eq:total_noise}, and changes as a function of $\theta$ no more than by a factor of two.}


{The above theoretical treatment gives a general idea of behavior of the FR noise in a birefringent medium. We see that the birefringence affects, in a certain way, behavior of the FR and ellipticity noises, but the most important result is that the total polarization noise is not suppressed in magnitude as it occurs with the regular magneto-optical effects.}

Let us consider now consequences of these results and what happens if we release some of the assumptions above.

{\it Discussion.} First of all, 
the main results of the treatment remain valid for arbitrary light propagation direction, with the dielectric constants along principal axes of the crystal being replaced by those along the normal-mode polarizations. Thus, by rotating the crystal, one can effectively change its birefringence varying in this way the wave vector $q$. This gives access to the parameters of spin system that may reveal themselves in its spatial correlations, like, e.g., the spin diffusion coefficients and the spin-orbit coupling constants. A similar idea was considered in~\cite{Kozlov2018a} where the wave vectors of the two waves differed by their directions rather than by magnitudes. 
The measurement of the spin noise at the wave vector $q$ is analogous to the measurement of the spin noise component at the Larmor frequency in the electron spin resonance~\cite{zapasskii-polar-anisotr99}.


Our theoretical treatment was performed for the case of a single homogeneously broadened optical resonance. The inhomogeneous broadening of the optical spectrum, as it follows from the above relations for $\theta=\pi/4$, generally leads only to 'isotropization' of the polarization noise on the Poincar\'e sphere (the arc at the point $A$ in Fig.~\ref{fig:poincare} turns into circle). If the broadening is much larger than the homogeneous width of the resonance, the noise intensities of the FR and ellipticity become equal and uncorrelated for any polarization direction of the incident light. Further we will demonstrate the experimental realization of this limit.

Since the polarization noise in a birefringent medium is practically insensitive to the magnitude and direction of the birefringence, we may conclude that this noise should not be affected by a spatially nonuniform birefringence. In other words, we can argue that spatial nonuniformity of the birefringent medium, does not hinder the spin-noise measurements. This result is very important for application of the spin noise spectroscopy to spatially inhomogeneous anisotropic media like multilayered structures or vitreous materials.


One more effect of linear birefringence on the measured polarization noise can be related to nonmonochromaticity of the probe light. After passing through the sample, this light becomes partially depolarized, but it preserves its $\xi_3$ Stokes parameter (provided that the medium is homogeneous). In this case, the ellipticity noise is suppressed (for $|\xi_3|\neq1$), while the FR noise intensity remains the same. This result is interesting, but does not have much practical sense for conventional laser sources with extremely narrow emission spectra.

Thus, we can conclude that the optical spin noise measurements are feasible for a broad class of birefringent media. Importantly, the gyration noise in the birefringent medium, does not reveal any pronounced dependence on the polarization plane azimuth.

\vspace{10pt}
{\it Experimental.} To verify basic results of our theoretical treatment, we have chosen two uniaxial crystals, CaWO$_4$ and LiYF$_4$, activated by Nd$^{3+}$ ions. 
An additional advantage of 
these crystals is that they contain tetragonal Nd$^{3+}$ centers aligned along the crystal axis, that are magnetically equivalent in external magnetic field~\cite{garrett-pmr-nd-cawo64,guedes-ident-re-imp02}. 
 
For optically anisotropic materials, the sample, 
generally, strongly affects polarization of the transmitted light. 
As a consequence, procedure of the spin-noise measurements, generally, needs to be modified. First, one should keep in mind that the FR noise, in a birefringent medium, is 
usually accompanied by the ellipticity noise (even in the region of transparency), 
with both of them being equally informative. 

\begin{figure*}[ht]
\includegraphics[width=\linewidth]{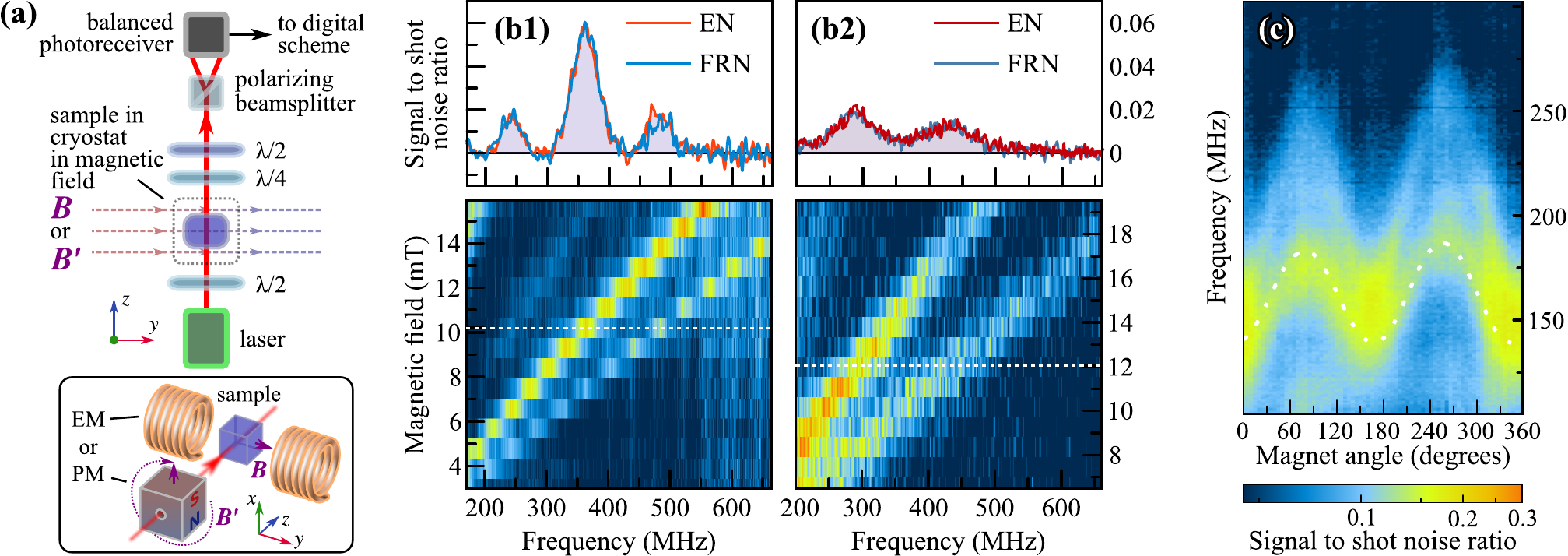}
\caption
{(a) Schematic of the experimental setup. The transverse magnetic field of variable magnitude ($B$) was created by the electromagnet (EM). Orientational measurements of the SN spectra were performed in the magnetic field ($B'$) of the permanent magnet (PM) rotating either around the laser beam, as shown in the inset, or around the axis normal to the laser beam (not shown). When measuring the ellipticity noise, the axes of the quarter-wave plate were aligned along the axes of the polarization ellipse of the transmitted light. When measuring the FR noise, the $\lambda$/4 plate was removed. Panel (b) shows magnetic-field dependences of the FR noise (lower pictures) and examples of particular records of the FR and ellipticity noise spectra (upper plots) for the Nd-doped CaWO$_4$ (b1) and LiYF$_4$ (b2) crystals. (c) Orientational dependence of the FR noise spectra of the CaWO$_4$:Nd$^{3+}$ crystal. Magnetic field created by the PM was rotated in the plane normal to the laser beam. Both resonances follow the same orientational pattern. The low-frequency resonance presumably corresponds to the known tetragonal neodymium center (accentuated with dotted line).}
\label{fig:exp}
\end{figure*}

{In homogeneously birefringent materials,} one should distinguish two main polarization schemes, with the probe light polarized along or at 45 degrees to the anisotropy axes of the crystal (as was illustrated above on the Poincar\'e sphere). In the first case, the light transmitted through the sample retains its linear polarization, and the measurements of the FR and ellipticity noise practically do not differ from those in isotropic materials. In the second case, polarization of the transmitted beam is controlled by phase retardation between the two normal modes, which may be arbitrary, and by their amplitudes, which may change due to dichroism of the crystal. As a result, angular position of the polarization ellipse of the light transmitted through the sample should be adjusted for each particular measurement (to balance the polarimetric detector). Similarly, for measuring the ellipticity noise, the axes of the quarter-wave plate, placed after the sample, should be aligned along the axes of the light-polarization ellipse, whose orientation also cannot be predicted in advance. Schematic of the optical arrangement with some additional explanations is presented in Fig.~\ref{fig:exp}(a). 

In the spatially-inhomogeneous birefringent samples, the above procedure cannot be easily realized. In this case, the polarization noise, as was already noted, retains its magnitude, but differentiation between the FR and ellipticity noise loses its sense. We will consider this situation in more detail elsewhere.

The plates of the crystals CaWO$_4$:Nd$^{3+}$ (1 at.\%), 1.54 mm thick, and LiYF$_4$:Nd$^{3+}$ ({approximately 0.5 at}.\%), 1.22 mm thick, were placed inside a cryostat at a temperature of 3 K. The quantity $qL$, for these samples, was $\sim 27$ (with the birefringence $\Delta n \approx 0.016$~\cite{bond-reflect-cryst65} for calcium tungstate and 32 (with the birefringence $\Delta n \approx$ 0.022)~\cite{barnes-ylf80} for LiYF$_4$, respectively, that well meets approximation of the above treatment ($qL \gg 1$). 
 Magnetic field on the sample was created either by an electromagnet, EM, or by a permanent magnet, PM (see the inset in Fig.~\ref{fig:exp}(a)). In the first case, magnetic field was directed horizontally, across the probe beam, and could be varied in magnitude. In the second, magnetic field of a fixed magnitude could be varied in direction (by rotating the PM \cite{kamenskii-re-sns2020}). 

As a light source, we used a frequency-stabilized ring-cavity Ti-sapphire laser with an extremely narrow emission spectrum needed to detect spin noise of RE ions in crystals \cite{kamenskii-re-sns2020}. The laser beam was tuned in resonance with the low-energy component of the transition $^4I_{9/2}$--$^4F_{3/2}$ of Nd$^{3+}$ ion characterized by a relatively small homogeneous width~\cite{han-spectroscopy-nd93}.
 
 
Our first measurements were performed in the simplest polarization scheme, with the light polarization aligned along an anisotropy axis of the crystal, in the magnetic field created by the electromagnet. In this case, the light beam, the magnetic field, and the optic axis of the crystal were perpendicular to each other. We have immediately found that, indeed, optical anisotropy of the crystal had no substantial effect upon the polarization noise signal, and the spin noise spectra of Nd$^{3+}$ ions could be detected in both crystals approximately with the same sensitivity as in the cubic CaF$_2$ crystal studied in~\cite{kamenskii-re-sns2020}.

Results of these measurements are shown in Fig.~\ref{fig:exp}(b,c). 
Panels (b1) and (b2) show magnetic-field dependences of the FR noise spectra (lower pictures) and examples of the FR and ellipticity noise spectra (upper plots) for the Nd-doped CaWO$_4$ and YLF crystals, respectively. Figure~\ref{fig:exp}(c) shows orientational dependence of the FR noise spectra of the CaWO$_4$-Nd$^{3+}$ crystal 
obtained in the magnetic field of the permanent magnet rotating in the plane normal to the light beam (as shown in Fig.~\ref{fig:exp}(a)). In this work, we did not mean to study in detail EPR spectra of the two samples. Still, the above results allowed us to ascribe the strongest peaks of the FR noise spectrum in the CaWO$_4$ and LiYF$_4$ crystal to the tetragonal Nd$^{3+}$ centers with $g$-tensor components $g_\|$~=~2.03, $g_\bot$~=~2.54)~\cite{garrett-pmr-nd-cawo64} and $g_\|$~=~1.987, $g_\bot$~=~2.554)~\cite{aminov-ylf90}, respectively. 

As seen from Fig.~\ref{fig:exp}(b) (upper plots), the FR and ellipticity noise signals, for the probe beam polarized along the optic axis of the crystal (at $\theta$ = 0$^o$), are the same in magnitude. This result well agrees with the theory. Our measurements performed at $\theta$ = 0$^\circ$, 45$^\circ$ and 90$^\circ$ have shown, however, that, within the experimental accuracy, the signals of the FR and ellipticity noise are universally equal to each other. {From the above theoretical treatment we can conclude that this situation is relevant for the strong inhomogeneous broadening. In addition, there are a few other possible alternatives: (i) The ratio of the FR and ellipticity noise intensities, at certain specific values of $(k_x-k_y)L$, may become weakly dependent on the angle $\theta$. (ii) There are real experimental parameters that were ignored in the above treatment like, e.g., effects of optical nonlinearity, contribution of the impurity-ion transition into birefringence of the host matrix, laser frequency jitter, etc. Altogether, 
what} is important from the viewpoint of this work is that the spin noise in anisotropic crystals can be detected as easily as in isotropic, and that, in these measurements, one should not even take care of the probe beam polarization.

{\it Conclusions.}
In this {Letter}, we show, both theoretically and experimentally, that the spatially uncorrelated fluctuations of magneto-optical activity, in contrast to the spatially uniform magneto-optical effects, are not affected essentially by linear birefringence of the medium and appear to be practically isotropic with respect to the probe beam polarization. {In the case of spatially correlated gyration noise, the detected noise signal may depend on the ratio of its correlation length and the length of polarization beats of the probe light (${1/q}$). This fact can be used for studying correlation properties of spin systems. An important prediction of this paper is that the FR noise is also unaffected by spatially inhomogeneous birefringence which may be useful for practical applications of the SNS to layered optically-anisotropic structures or glasses. We believe that results of this work considerably strengthen potential of the spin noise spectroscopy. }

\begin{acknowledgments}
  The authors highly appreciate the Russian Science Foundation Grant No. 21-72-10021, within which framework the main paper idea statement and the fulfillment of experimental research was done. The development of the microscopic theory of the Faraday rotation noise in homogeneous birefringent media by D. S. S. was partially financially supported by the RF President Grant No. MK-5158.2021.1.2, the Foundation for the Advancement of Theoretical Physics and Mathematics ``BASIS,'' and Russian Foundation for Basic Research Grant No. 19-52-12038.
\end{acknowledgments}

\bibliography{biblio}
\bibliographystyle{apsrev4-2}
 
\end{document}